\documentclass[aps,pra,reprint,floatfix]{revtex4-1}
\usepackage{amsmath}
\usepackage{amssymb}
\usepackage{multirow}
\usepackage{graphicx}
\usepackage{babel}
\usepackage{hyperref}
\usepackage{natbib}
\usepackage{siunitx}
\usepackage[T1]{fontenc}
\usepackage{color}
\usepackage{epstopdf}

\begin{document}

\title{Effect of interfacial intermixing on spin-orbit torque in Co/Pt bilayers}

\author{G. G. Baez Flores and K. D. Belashchenko}

\affiliation{Department of Physics and Astronomy and Nebraska Center for Materials and Nanoscience, University of Nebraska-Lincoln, Lincoln, Nebraska 68588, USA}

\date{\today}
\begin{abstract}
Using the first-principles non-equilibrium Green's function technique with supercell disorder averaging, we study the influence of interfacial intermixing on the spin-orbit torque in Co$\mid$Pt bilayers. Intermixing is modeled by inserting one or more monolayers of a disordered CoPt alloy between Co and Pt. Dampinglike torque is moderately enhanced by interfacial intermixing, while the fieldlike torque, which is small for abrupt interfaces, is strongly enhanced and becomes comparable to the dampinglike torque. The enhancement of the fieldlike torque is attributed to the interface between Co and the intermixed region. The planar Hall-like torque increases with intermixing but remains relatively small. The behavior of the torques is similar for bilayers with (111) and (001)-oriented interfaces. Strong dependence of the fieldlike torque on intermixing could provide a way to tune the fieldlike-to-dampinglike torque ratio by interface engineering.
\end{abstract}
\maketitle

\section{Introduction}

Current-induced spin-orbit torque (SOT) in ferromagnet/heavy-metal (FM/HM) bilayers enables the manipulation and switching of the magnetization in magnetic memories, nano-oscillators, and other spintronic devices \cite{Miron2011,Liu2012,Manchon}. Optimizing the materials and interfaces for more efficient generation of SOT is an important goal for device applications \cite{Manchon,zhu2021maximizing,Song-review,SOT-roadmap}.

SOT is a manifestation of charge-to-spin conversion arising thanks to spin-orbit coupling (SOC) through a variety of microscopic mechanisms. One mechanism involves the spin-Hall current \cite{Sinova-SHE} generated in the HM layer by the electric field $\mathbf{E}$ and absorbed by the FM. This mechanism results in the dominant dampinglike (DL) SOT with the $\mathbf{m}\times(\mathbf{y}\times\mathbf{m})$ angular dependence and a relatively small fieldlike (FL) SOT proportional to $\mathbf{m}\times\mathbf{y}$ \cite{SOTGmix}. Here $\mathbf{m}$ is the unit vector parallel to the magnetization, and $\mathbf{y}=\mathbf{z}\times\hat{\mathbf{E}}$ is the polarization axis of the spin Hall current flowing along the interface normal $\mathbf{z}$. Another mechanism involves the inverse spin-galvanic effect (ISGE) \cite{AronovGeller1989JETPL,Edelstein1990,Ganichev} generating a nonequilibrium spin accumulation at the interface, which contributes primarily to the FL SOT \cite{SOTGmix}.

Material-specific prediction of SOT and the interpretation of experimental data are not straightforward. Even in the large-thickness limit of the standard spin-diffusion model, DL SOT in the FM layer depends not only on the spin-Hall conductivity of the HM layer, but also on the ratio of the spin-mixing conductance of the interface and the spin conductance of the HM layer \cite{SOTGmix}. For the commonly used FM/Pt interfaces this ratio is not large, and the so-called spin backflow reduces DL SOT at least by half \cite{zhu2021maximizing}. In addition, SOC at the interface can partially convert spin current into angular momentum of the lattice and modify the boundary conditions for transport.

To complicate things even further, interfaces can serve as sources of spin currents and torques \cite{Amin-APR}. Thus, any modification of the interface can change the SOT acting on the FM layer, which is both a challenge and an opportunity for interface engineering.

First-principles calculations show that interfacial intermixing changes the spin-dependent interface resistances \cite{Xia2001}, real and imaginary parts of the spin-mixing conductance \cite{SOT_FMNM}, and interfacial spin relaxation (spin memory loss) \cite{Gupta,GGBF}. Although spin memory loss for longitudinal spin current is not directly related to SOT \cite{GGBF}, intermixing should also affect SOC-induced absorption of spin current by the lattice. Experimentally, the degree of interfacial disorder and intermixing is difficult to characterize and control \cite{Chiang1998,Zambano2002,Cyrille2000,Bass-JMMM}. It was found that an insertion of a 0.6-nm Co-Pt alloy interlayer between Co and Pt increases the DL torquance by approximately 30\% \cite{Zhu2019}.

The nonequilibrium Green's function (NEGF) approach \cite{datta1997book,Nikolic2018} with supercell disorder averaging based on the first-principles electronic structure is ideally suited for materials-specific studies of current-induced SOT. This technique was implemented \cite{Faleev2005} within the tight-binding linear muffin-tin orbital method (TB-LMTO) \cite{Turek} in the Questaal package \cite{QUESTAAL} and adapted for the studies of SOT \cite{kdbelSOT,KDB2}. The magnitude of the DL SOT calculated for Co$\mid$Pt (001) bilayers with abrupt interfaces \cite{kdbelSOT,KDB2} was in reasonable agreement with experiments. The asymptotic value of the spin-torque efficiency $\xi^E_\mathrm{DL}$ at large Pt thickness was about \SI{2.9e5}{\per\ohm\per\meter} for a Co$\mid$Pt bilayer with 4 monolayers (ML) of Co and resistivity of \SI{27}{\micro\ohm\cm}, where $\xi^E_\mathrm{DL}=(2e/\hbar) T_\mathrm{DL}/E$ and $T_\mathrm{DL}$ is the areal density of DL SOT. A similar value of $\xi^E_\mathrm{DL}$ was reported, for example, in Ref.\ \onlinecite{Nguyen2016}. NEGF calculations for Co$\mid$Pt (001) bilayers also showed a moderate FL SOT (10-15\% of DL SOT in the asymptotic limit) and a small but important planar-Hall-like SOT which contributes a term proportional to $m_xm_z$ to magnetization damping. This latter term has been observed in spin-torque ferromagnetic resonance measurements \cite{Safranski2019}.

In this paper, we study the effects of interfacial intermixing on SOT in Co$\mid$Pt bilayers with (111) and (001) surface orientations using the NEGF approach. We find that intermixing results in a moderate enhancement of DL SOT, in agreement with the measurements of Ref.\  \onlinecite{Zhu2019}, and a strong enhancement of FL SOT, which we attribute to the interface between Co and the intermixed region. This strong dependence of FL SOT on intermixing could be used to tune the FL/DL torque ratio for optimal device performance.

\section{Computational details}

We focus on Co$\mid$Pt bilayers with the (111) growth direction, but (001) systems are also used for comparison. Interfacial intermixing is modeled by inserting one or more disordered layers of Co$_{0.5}$Pt$_{0.5}$ (which we denote as M) between Co and Pt. For further insight, we also consider Co$\mid$M and M$\mid$Pt structures. The systems are listed in Table \ref{tab_result} where the layer thicknesses in monolayers are given in brackets. The outer surfaces in each structure are separated by a vacuum region.

\begin{table}[htb]
    \centering
    \begin{tabular}{|r|l|c|c|c|}
        \hline
         \# & System & $\xi^E_\mathrm{DL}$  & $\xi^E_\mathrm{FL}$ & $\xi^E_\mathrm{PHL}$ \\
        \hline
        1 & Co(6)$\mid$Pt(6) (001) & $1.40$ & $-0.08$ & $0.14$\\
        \hline
        2 & Co(6)$\mid$Pt(6) & $1.20$ & $0.28$&$0.14$\\
        \hline
        3 & Co(6)$\mid$Pt(12) & $1.37$&$0.11$&$0.11$\\
        \hline
        4 & Co(6)$\mid$Pt(24) &$1.83$&$0.05$&$0.15$\\
        \hline
        5 & Co(6)$\mid$M$\mid$Pt(6) & $1.35$&$-0.45$&$0.14$\\
        \hline
        6 & Co(6)$\mid$M$\mid$Pt(12) &$1.68$&$-0.50$&$0.13$\\
        \hline
        7 & Co(6)$\mid$M$\mid$Pt(24) &$1.84$&$-0.42$&$0.16$\\
        \hline
        8 & Co(6)$\mid$M(2)$\mid$Pt(6) &$1.48$&$-0.50$&$0.15$\\
        \hline
        9 & Co(6)$\mid$M(3)$\mid$Pt(6) &$1.61$&$-0.71$&$0.20$\\
        \hline
        10 & Co(6)$\mid$M(6)$\mid$Pt(6) & $1.64$&$-0.97$&$0.27$\\
        \hline
        11 & Co(6)$\mid$M(3)$\mid$Pt(12) &$1.96$&$-0.83$&$0.26$\\
        \hline
        12 & Co(6)$\mid$M(3)$\mid$Pt(24) & $2.18$&$-0.89$&$0.20$\\
        \hline
        13 & Co(6)$\mid$M$'$(3)$\mid$Pt(6) & $1.47$&$-0.74$&$0.19$\\
        \hline
        14 & Co(6)$\mid$M(3)$\mid$Pt(6) (001) & $1.57$&$-0.99$&$0.20$\\
        \hline
        15 & Co(6)$\mid$M(6) &$0.21$&$-1.15$&$0.20$\\
        \hline
        16 & M(3)$\mid$Pt(6) & $1.61$&$-0.30$&$0.04$\\
        \hline
        17 & M(6)$\mid$Pt(6) &$1.68$&$-0.09$&$0.07$\\
        \hline
        18 & O(1)|Co(6) & $-0.07$ & $0.25$ & $0$ \\
        \hline
    \end{tabular}
    \caption{SOT efficiencies $\xi^E_q$ in units of $10^5$ \SI{}{\per\ohm\per\meter}. Layers are listed in the order of decreasing $z$ coordinate, and numbers in brackets indicate layer thicknesses in monolayers. Disordered Co$_{0.5}$Pt$_{0.5}$ layers are denoted as M. M$^\prime$(3) denotes three layers with concentrations of 75, 50, and 25\%. The growth direction is (111) if not noted. Typical error bars from disorder sampling are \SI{0.1e5}{\per\ohm\per\meter} for $\xi^E_\mathrm{DL}$ and $\xi^E_\mathrm{FL}$ and \SI{0.05e5}{\per\ohm\per\meter} for $\xi^E_\mathrm{PHL}$.}
    \label{tab_result}
\end{table}

Structural optimizations were performed using the Vienna Ab Initio Simulation Package (VASP) \cite{VASP1,VASP2,VASP3,VASP4} with PAW pseudopotentials \cite{Blochl,PAW} and the generalized gradient approximation \cite{PBE}. The Co(6)$\mid$Pt(6) system was fully optimized. The interlayer distances vary from 1.85 to \SI{1.98}{\angstrom} for Co and from 2.42 to \SI{2.46}{\angstrom} for Pt; the Co-Pt interlayer distance is \SI{2.14}{\angstrom} and the nearest Co-Pt distance is \SI{2.64}{\angstrom}. To build the structures with intermixed layers, we started with a fully optimized Co(6)$\mid$M(1)$\mid$Pt(6) $2\times1$ lateral supercell. The buckling of the M layer and the lateral shifts of the atoms are insignificant in this relaxed supercell. In the TB-LMTO calculations, the Co and Pt atoms in each intermixed monolayer were assumed to lie in the same plane, and the lateral shifts from the high-symmetry sites were neglected. The relaxed Co(6)$\mid$Pt(6) and Co(6)$\mid$M(1)$\mid$Pt(6) structures both have an in-plane lattice parameter of $\SI{2.67}{\angstrom}$ which was used for all structures. In VASP calculations the vacuum region between the two outer surfaces was 1 nm thick, and in TB-LMTO calculations it was represented by 4--6 monolayers of empty spheres, which amounts to 0.87--1.3 nm.

Systems with more than six Pt monolayers were constructed by inserting additional Pt monolayers in the middle of the Pt layer with the same interlayer distance. Co$\mid$M and M$\mid$Pt bilayers were obtained by truncating the corresponding Co$\mid$M$\mid$Pt structure. Systems with two or more M layers were obtained by inserting additional M layers with the M-M interlayer distance of \SI{2.20}{\angstrom}, which gives the same volume per atom in the epitaxially constrained disordered Co-Pt alloy as in the fully relaxed L1$_1$-ordered CoPt structure. The Co(6)$\mid$M(3)$\mid$Pt(6) system with (001) interfaces was built using the ideal face-centered cubic lattice as in Refs.\ \onlinecite{kdbelSOT,KDB2}. 

The spin moments of Co and Pt atoms in three representative structures are shown in Table \ref{tab:magmom}. Those of the Co atoms are slightly enhanced near the interfaces and in the M layers compared to bulk Co. Pt atoms inside the M layers have sizeable spin moments of about 0.3 $\mu_B$, and there is a notable proximity effect in the Pt layers nearest to the ferromagnetic layers.

\begin{table}[htb]
    \centering
    \begin{tabular}{|c|c|c|c|}
    \hline
       Atom & Co|Pt & Co|M|Pt & Co|M(3)|Pt  \\
    \hline
         Co$_1$ & 1.79 & 1.80 & 1.81\\
    \hline
         Co$_2$ & 1.69 & 1.69 & 1.70\\
    \hline
         Co$_3$ & 1.72 & 1.72 & 1.73 \\
    \hline
         Co$_4$ & 1.71 & 1.72 & 1.72\\
        \hline
         Co$_5$ & 1.71 & 1.72 &  1.73\\
         \hline
         Co$_6$ & 1.77 & 1.75 & 1.74 \\
         \hline
         M$_1$   & --- & 1.79, 0.29 & 1.76, 0.33\\ 
         \hline
         M$_2$   & --- & --- &  1.80, 0.31\\
         \hline
         M$_3$   & --- & --- & 1.89, 0.22\\
         \hline
         Pt$_1$ & 0.17 & 0.10 & 0.09 \\
         \hline
         Pt$_2$ & 0.03 & 0.03 & 0.03\\ 
         \hline
    \end{tabular}
    \caption{Magnetic moments of Co and Pt in representative structures ($\mu_B$).}
    \label{tab:magmom}
\end{table}

SOT was calculated using the NEGF approach \cite{Faleev2005,kdbelSOT,KDB2} within the TB-LMTO method in the atomic sphere approximation \cite{Turek,QUESTAAL}. The sites in the disordered layers were randomly occupied with Co and Pt atoms, usually with a 50\% concentration. The charge and spin densities were obtained using self-consistent calculations treating the disordered layers within the coherent potential approximation (CPA) \cite{Turek,QUESTAAL}. The radii of the Co and Pt atomic spheres were adjusted so that they have equal charges in each intermixed layer. This ensures that substitutional disorder does not generate random Madelung potentials. The leads had the same structure as the active region, except that the intermixed layers were fully occupied by Co atoms.

The active region was taken to be 70 atomic layers in length ($\SI{16.2}{\nano\meter}$) and 4 in width ($\SI{0.48}{\nano\meter}$). The current flow was along the [120] crystallographic direction. Uniformly distributed Anderson disorder potential with the maximum magnitude of \SI{1.09}{\eV} was used, which results in the resistivities of \SI{27}{} and \SI{29}{\micro\ohm\centi\meter} for Co$\mid$Pt and Co$\mid$M$\mid$Pt bilayers, respectively; these are typical values observed in experiments \cite{Resistivity1,Resistivity2,Safranski2019}. 25 disorder configurations were used for each system, and five monolayers near each lead were excluded from the averaging. We have checked that edge effects near the leads are negligible in the calculated SOT.

Site-resolved torquances are defined as $\boldsymbol{\tau}_i(\mathbf{m})=\mathbf{T}_i(\mathbf{m})/E$, where $\mathbf{T}_i(\mathbf{m})$ is the torque on site $i$ for the given orientation of the magnetization unit vector $\mathbf{m}$. These quantities were calculated for 32 orientations of $\mathbf{m}$ and projected on the orthonormal basis set of real vector spherical harmonics $\boldsymbol{Z}^{(\nu)}_{lm}$ (VSH) \cite{KDB2}.

The electric field $E$ in the diffusive embedded region is determined as $E=VGdR/dL$, where $V$ is the voltage drop, $G$ the Landauer-B\"uttiker conductance of the supercell embedded between the two leads, $R=1/G$, and $L$ the length of the supercell. This expression takes into account that a relatively small portion of the voltage drop occurs at the edges of the supercell. In practice, the NEGF technique yields the Fermi-surface contribution to the torque in the form of dimensionless linear-response site-resolved quantities $\mathbf{t}_i=e^{-1}d\mathbf{T}_i/dV$. The torquances are then obtained as $\boldsymbol{\tau}_{i}=e \mathbf{t}_{i}R(dR/dL)^{-1}$ and expressed in units of $ea_0$, where $a_0$ is the Bohr radius.

In the VSH expansion, all terms with $m=(-1)^l$ are allowed in any bilayer, including the axially-symmetric case ($C_{\infty v}$ symmetry) \cite{KDB2}. The lowest-order terms are the DL $\boldsymbol{Z}^{(1)}_{1,-1}=a\mathbf{m}\times(\mathbf{y}\times\mathbf{m})$ and FL $\boldsymbol{Z}^{(2)}_{1,-1}=a\mathbf{m}\times\mathbf{y}$ harmonics, where $a=\sqrt{3/8\pi}$ is the normalization factor. The planar Hall-like (PHL) $\boldsymbol{Z}^{(1)}_{2,1}$ harmonic \cite{Safranski2019,kdbelSOT,KDB2,PHLdef} is also reliably identified. The $C_{3v}$ symmetry of the (111) interface allows additional terms with $m=2(-1)^{l+1}$ and $m=4(-1)^{l+1}$, such as $\mathbf{Z}^{(\nu)}_{2,-2}$, $\mathbf{Z}^{(\nu)}_{3,2}$, etc., but they do not rise above the disorder sampling noise in our calculations.

We report the torquances using the common normalization factor that gives the usual definitions of the DL and FL components:
\begin{align}
\boldsymbol{\tau}_i(\mathbf{m}) & =\tau^i_{\mathrm{DL}}\mathbf{m}\times(\mathbf{y}\times\mathbf{m}) + \tau^i_{\mathrm{FL}}\mathbf{m}\times\mathbf{y}\nonumber\\
& +\tau^i_{\mathrm{PHL}}a^{-1}\mathbf{Z}^{(1)}_{2,1}\,
\label{def}
\end{align}

With this definition, positive $\tau^i_{\mathrm{FL}}$ corresponds to a current-induced effective field that is opposite in sign to the Oersted field in a typical experiment.
To facilitate comparison with experimental data, torquances integrated over the thickness of the bilayer are given as torque efficiencies $\xi^E_{q}=(2e/\hbar)\tau_{q}/A$ where $\tau_{q}=\sum_i\tau^i_q$ is the total torquance of type $q$ (i.e., DL, FL, or PHL) per area $A$.

\section{Results and discussion}

Table \ref{tab_result} lists the computed torquances. First, we note that $\xi^E_\mathrm{DL}$ in Co(6)$\mid$Pt(6) is quite similar for (001) and (111) interface orientations. This is expected of a DL SOT coming from the spin-Hall effect in Pt if the two interfaces have similar spin-mixing conductances and spin relaxation. However, a large effective interfacial contribution was found \cite{KDB2} for the Co/Pt (001) interface, and the present results for Co(6)$\mid$Pt(N) at $N=6$, 12, and 24 (Systems 2-4 in Table \ref{tab_result}) suggest a similar feature for (111)-oriented bilayers. The similarity of the DL SOT for (001) and (111) interfaces suggests that the effective interfacial contributions are also similar. The PHL SOT is also similar for (001) and (111) interfaces.

The FL SOT is positive and amounts to approximately 23\% of the DL SOT in Co(6)$\mid$Pt(6) (111); it declines with increasing thickness of Pt (Systems 2-4). For the (001) interface the FL SOT is small in Co(6)$\mid$Pt(6) (System 1). It increases at larger Pt thicknesses while still remaining small compared to DL SOT \cite{KDB2}. Thus, FL SOT exhibits some variation in Co$\mid$Pt bilayers with abrupt (001) and (111) interfaces but remains rather small compared to the DL SOT.

Now we turn to systems with interfacial intermixing. The addition of a monolayer of disordered Co$_{0.5}$Pt$_{0.5}$ alloy (labeled as M; Systems 5-7 in Table \ref{tab_result}) between pure Co and Pt layers leads to a marginal enhancement of the DL and PHL SOT. On the other hand, the FL SOT becomes negative and rather large. In principle, intermixing can enhance FL SOT by increasing the imaginary part of the spin-mixing conductance. Such an increase was found in calculations for the Co/Cu (111) interface \cite{SOT_FMNM}, although that increased imaginary part was still less than 6\% of the real part. However, if the enhancement of FL SOT by intermixing were due to this mechanism, we would expect FL SOT to increase with increasing thickness of Pt (Systems 5-7) in proportion to the increasing DL SOT. The calculated FL SOT does not exhibit such a trend, suggesting that its enhancement by intermixing is rather due to the modification of the interfacial inverse spin-galvanic effect.

Stronger intermixing is modeled by adding more intermixed layers in Co$\mid$Pt (111) bilayers (Systems 8-12 in Table \ref{tab_result}). We also consider a bilayer with a gradual transition from Co to Pt through 3 intermixed layers with concentrations of 75, 50, and 25\% (System 13), which mimics the system with artificial intermixing studied in Ref.\ \onlinecite{Zhu2019}, as well as a Co(6)$\mid$M(3)$\mid$Pt(6) (001) bilayer (System 14).

In Co$\mid$Pt (111) bilayers $\xi^E_\mathrm{DL}$ clearly increases with increasing intermixing. In the Co(6)$\mid$M($N$)$\mid$Pt(6) (111) sequence it appears to saturate at $N=3$ at a value that is enhanced by about 35\% compared to Co(6)$\mid$Pt(6) (System 2). Comparing Co(6)$\mid$M(3)$\mid$Pt($N$) and Co(6)$\mid$Pt($N$) bilayers, we find that intermixing consistently enhances $\xi^E_\mathrm{DL}$ at all Pt thicknesses (6, 12, or 24 layers). The value of $\xi^E_\mathrm{DL}$ in System 13 with the concentration gradient through 3 intermixed layers is similar to System 8 with 2 fully intermixed layers. $\xi^E_\mathrm{DL}$ is also larger in Co(6)$\mid$M(3)$\mid$Pt(6) (001) (System 14) compared to Co(6)$\mid$Pt(6) (001) (System 2), but only by about 13\%, which is considerably less compared to (111)-oriented bilayers.
Overall, the moderate enhancement of $\xi^E_\mathrm{DL}$ found in our calculations with intermixing is in qualitative agreement with experiment \cite{Zhu2019}, which found a 30\% enhancement of $\xi^E_\mathrm{DL}$ for Co$\mid$Pt bilayers with a 0.6-nm insertion of a Co-Pt alloy with a profile mimicked by our System 13.

Intermixing results in a substantial increase of $\xi^E_\mathrm{PHL}$, which is almost doubled in Co(6)$\mid$M(6)$\mid$Pt(6) (System 10) compared to Co(6)$\mid$Pt(6) (System 2). We also note that $\xi^E_\mathrm{PHL}$ is the same for (001) and (111) oriented bilayers both with and without intermixing (Systems 1 and 2; 9 and 14).

The most significant effect of intermixing is the emergence of a large negative $\xi^E_\mathrm{FL}$. In System 10 with 6 intermixed layers it is nearly $10^5$ \SI{}{\per\ohm\per\meter} in magnitude, or almost 60\% of $\xi^E_\mathrm{DL}$. The magnitude of $\xi^E_\mathrm{FL}$ is almost the same in Systems 9 and 13 which differ only in the concentration profile of the intermixed layer, and it is even larger in the (001)-oriented Co(6)$\mid$M(3)$\mid$Pt(6) bilayer (System 14).

Sizeable \emph{positive} FL SOT ($\tau_\mathrm{FL}>0$ which acts against the Oersted field) was reported in some measurements for Co$\mid$Pt \cite{Garello2013} and CoFe$\mid$Pt \cite{Pai2015} bilayers, as well as in Co \cite{Baumgartner2017} and (Co$\mid$Ni)$_N$ \cite{Figueiredo2021} disks grown on Pt. Other measurements show very small FL SOT \cite{Zhu2019}. To our knowledge, the origin of this variation is unknown.
Positive FL SOT was also found \cite{Freimuth2014} in linear-response calculations for the O(1)$\mid$Co(3)$\mid$Pt(10) system where the Co layer is capped with a monolayer of oxygen. We have similarly  
found a moderate positive $\xi^E_\mathrm{FL}=\SI{0.25e5}{\per\ohm\per\meter}$ in a O(1)$\mid$Co(6) system (System 18 in Table \ref{tab_result}), confirming that oxidation of the top surface tends to induce positive FL SOT. This oxidation-induced FL SOT is opposite in sign to the FL SOT induced in our calculations by substitutional intermixing at the Co$\mid$Pt interface.

Systems with several intermixed layers at the interface can be viewed as having three bulk layers (Co, M, and Pt) with two interfaces (Co$\mid$M and M$\mid$Pt) between them. Therefore, we also considered Co$\mid$M and M$\mid$Pt bilayers (Systems 15-17 in Table \ref{tab_result}). The results allow us to make the following three observations.

(1) $\xi^E_\mathrm{DL}$ is similarly large in all Co$\mid$Pt, Co$\mid$M$\mid$Pt, and M$\mid$Pt systems, but it is reduced by a factor of 8 in Co$\mid$M. This is consistent with DL SOT coming primarily from the Pt layer.

(2) $\xi^E_\mathrm{FL}$ in Co(6)$\mid$M(6) is larger than in Co(6)$\mid$M(6)$\mid$Pt(6), while in the M$\mid$Pt systems it is much weaker. This suggests that strong FL SOT in bilayers with intermixed interfaces originates from the inverse spin-galvanic effect at the Co$\mid$M interface. This behavior rules out the spin-Hall effect in Pt as the dominant origin of the FL SOT in these systems. The largest $\xi^E_\mathrm{FL}$ occurs in Co(6)$\mid$M(6) (System 15) where $\xi^E_\mathrm{DL}$ is the smallest, suggesting that the small contribution from the spin-Hall effect is opposite in sign to that from the inverse spin-galvanic effect.

(3) $\xi^E_\mathrm{PHL}$ is small in M$\mid$Pt (Systems 15 and 16), but in Co(6)$\mid$M(6) (System 15) it is similar to Co$\mid$M$\mid$Pt systems. Given that $\xi^E_\mathrm{PHL}$ is not much smaller in Co$\mid$Pt without intermixing (where $\xi^E_\mathrm{FL}$ is small and has an opposite sign), we conclude that $\xi^E_\mathrm{PHL}$ does not correlate well with either $\xi^E_\mathrm{DL}$ or $\xi^E_\mathrm{FL}$.

\begin{figure}[h!]
    \centering
    \includegraphics[width=0.95\linewidth]{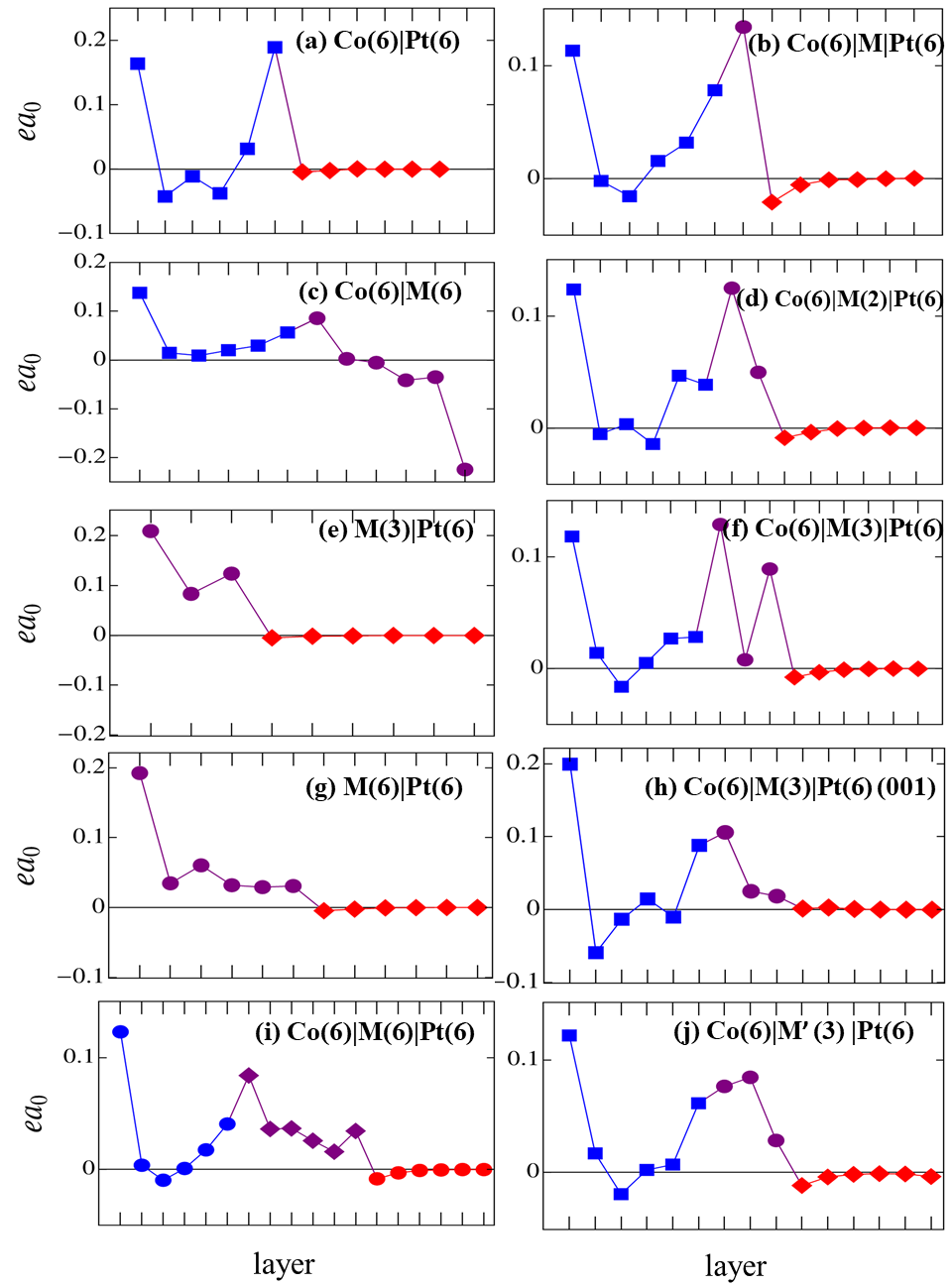}
    \caption{Layer-resolved DL torquance $\tau^i_\mathrm{DL}$ for the systems shown in the legends. Blue squares, red diamonds, and purple circles show monolayers of Co, Pt, and M=Co$_{0.5}$Pt$_{0.5}$, respectively.}
    \label{fig:atom_dl}
\end{figure}

\begin{figure}[h!]
    \centering
    \includegraphics[width=0.95\linewidth]{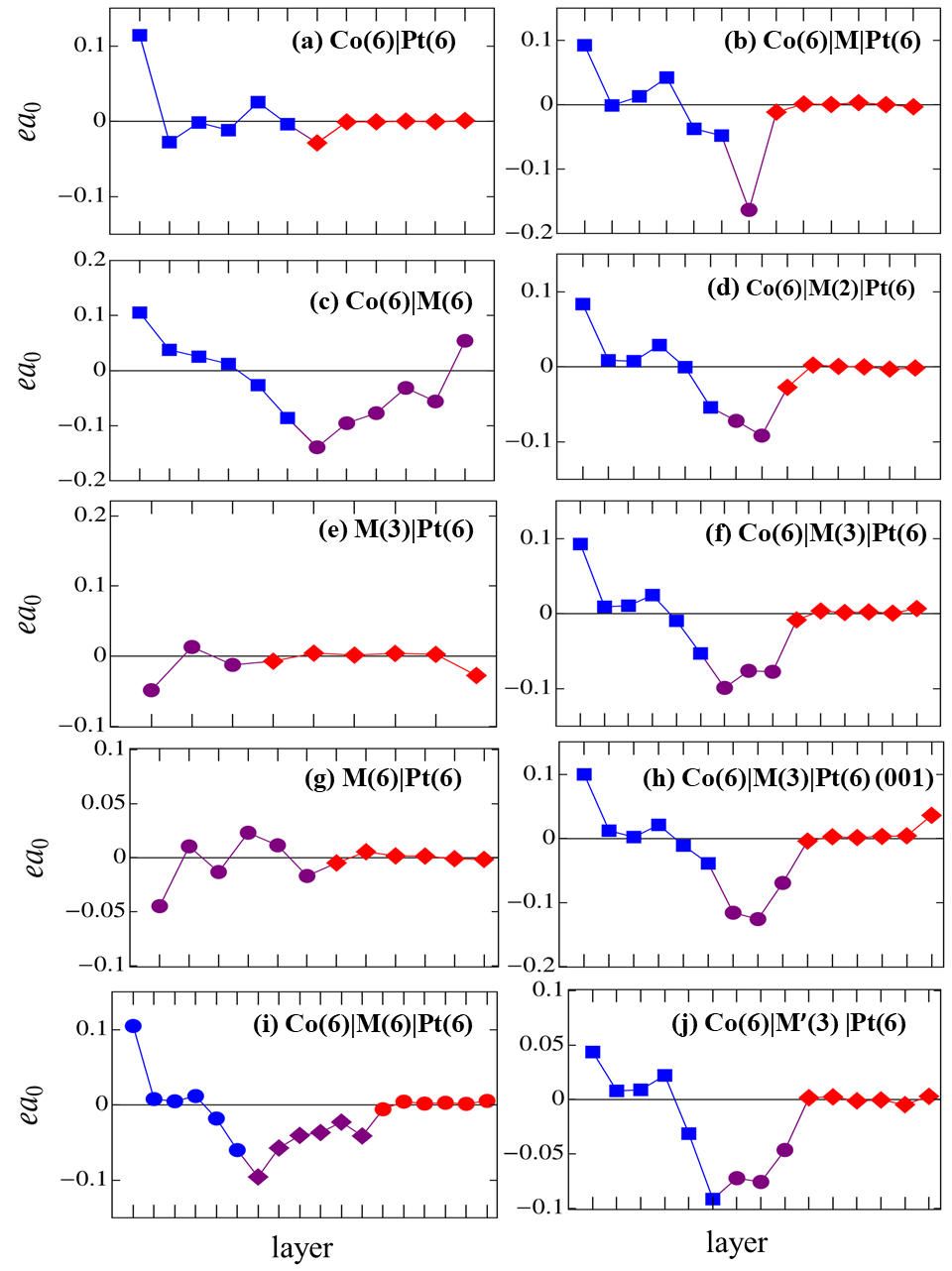}
    \caption{Same as in Fig.\ \ref{fig:atom_dl} but for the FL torquance $\tau^i_\mathrm{FL}$.}
    \label{fig:atom_fl}
\end{figure}

Additional detail can be obtained from the site-resolved torquances $\tau^i_\mathrm{DL}$ and $\tau^i_\mathrm{FL}$, which are shown for selected systems in Fig.\ \ref{fig:atom_dl} and \ref{fig:atom_fl}, respectively. Similar to Co$\mid$Pt (001) bilayers \cite{kdbelSOT}, we see a large DL SOT on the (111) free surface of Co. The DL SOT at the free surface of the Co$_{0.5}$Pt$_{0.5}$ alloy [panels (e) and (g)] is even slightly larger than that. Such ``anomalous SOT'' has been detected experimentally via the tilting of the magnetization at the surfaces of a ferromagnetic film \cite{Fan-ASOT}.

In an inversion-symmetric ferromagnetic layer with similar surfaces, the SOT at the two surfaces exactly cancel each other. Nearly perfect cancellation was found \cite{kdbelSOT} for a Co(6)$\mid$Pt(6) bilayer with spin-orbit coupling switched off on all Pt atoms, even though the two surfaces are different in that case. Strong cancellation was also observed in measurements of single ferromagnetic films with various capping layers \cite{Fan-ASOT}.
As noted above, $\xi^E_\mathrm{DL}$ is fairly small in the Co(6)$\mid$M(6) system. The site-resolved torquances shown in Fig.\ \ref{fig:atom_dl}(c) have opposite signs at the free surfaces of Co and Co$_{0.5}$Pt$_{0.5}$, and there is also positive torque localized at the Co$\mid$M interface. The surface and interfacial contributions may be due to the absorption of the spin-Hall currents generated in the ferromagnetic layers \cite{Amin-SHFM}, which suggests a possible explanation for the near-cancellation of the total DL SOT.

All other systems shown in Fig.\ \ref{fig:atom_dl} have a layer of Pt, which generates positive DL SOT at the FM$\mid$Pt interface (where FM stands for Co or M). This ``conventional'' DL SOT overcomes the negative self-generated SOT that would be seen at that interface in the absence of Pt [as in Fig. \ref{fig:atom_dl}(c)], and it extends deeper into the FM layer.

Figure \ref{fig:atom_fl} shows a common pattern in the FL SOT for all systems containing a Co$\mid$M interface: large negative FL SOT on the interfacial M layer, the adjacent Co layer, and a few additional M layers, if present. This feature is consistent with the above observation that large FL SOT develops in systems with a Co$\mid$M interface.

\section{Conclusions}

We studied the effects of interfacial intermixing on SOT in Co$\mid$Pt (111) and (001) bilayers using the TB-LMTO-NEGF technique with explicit supercell averaging over disorder configurations.
Intermixing increases the DL torque efficiency $\xi^E_\mathrm{DL}$ by 10-35\% depending on the growth direction, in qualitative agreement with the experimental finding of Ref.\ \onlinecite{Zhu2019}. Intermixing also induces a large negative FL SOT for both (111) and (001) interfaces, which is opposite in sign to the FL SOT found in some experiments with oxide-capped Co$\mid$Pt bilayers and in calculations where the Co layer is capped with a monolayer of oxygen. This FL SOT persists in Co$\mid$M but is absent in M$\mid$Pt bilayers (where M stands for disordered Co$_{0.5}$Pt$_{0.5}$), showing that the strong FL SOT originates at the Co$\mid$M interface. 
Strong dependence of the FL/DL torque ratio on the thickness of the intermixed region in Co$\mid$Pt bilayers could provide a way to control this ratio by interface engineering.
The planar Hall-like torque is also enhanced by intermixing but remains relatively small compared to DL and FL torques. 

\begin{acknowledgments}
We are grateful to Xin Fan for useful comments on the manuscript. This work was supported by the National Science Foundation through Grant No. DMR-1916275. Calculations were performed utilizing the Holland Computing Center of the University of Nebraska, which receives support from the Nebraska Research Initiative.
\end{acknowledgments}

\bibliography{bibfile}
\end{document}